\documentclass[a4paper,12pt]{article}
\usepackage{amssymb,amsthm,latexsym}
\newcommand{\Med}[1]{\left\langle #1 \right\rangle}

\newtheorem{theorem}{Theorem}

\title{The infinite volume limit in generalized mean field disordered 
models}

\author{
Francesco Guerra\footnote{\
e-mail: {\tt francesco.guerra@roma1.infn.it}} \\
{\small {\itshape Dipartimento di Fisica, Universit\`a di Roma `La 
Sapienza'}}
\\
{\small {\itshape and INFN, Sezione di Roma, Piazzale A. Moro 2, 00185 
Roma, 
Italy}}\\
Fabio Lucio Toninelli\footnote{\ 
e-mail: {\tt f.toninelli@sns.it}} \\
{\small {\itshape Scuola Normale Superiore, Piazza dei Cavalieri 7, 56126 
Pisa,
Italy}}\\
{\small {\itshape and Istituto Nazionale di Fisica Nucleare, Sezione di 
Pisa}}
} 

\date{\today}

\begin{document}

\maketitle

\begin{abstract}
We generalize the strategy we recently introduced to prove the existence 
of 
the thermodynamic limit for the Sherrington-Kirkpatrick 
and $p$-spin models,
to a wider class of mean field spin glass systems, including models with 
multi-component and non-Ising type spins, mean field spin glasses with an 
additional Curie-Weiss interaction, and systems consisting 
of several replicas of the
spin glass model, where replicas are coupled with terms depending on the 
mutual overlaps.
\end{abstract}

\newpage

\section{Introduction}

In a recent paper \cite{limterm}, we introduced a simple interpolation 
method,
which allows to prove the existence of the thermodynamic limit for the 
quenched
average of the free energy and ground state energy per site, for a wide 
class
of mean field spin glass models \cite{MPV}. This class includes, for 
instance,
the well known 
Sherrington-Kirkpatrick (SK) model \cite{sk, sk2} and Derrida's $p$-spin 
model \cite{derrida, gross, gardner}, for even $p$.
Moreover, we proved almost sure convergence, with respect to the quenched 
disorder, without taking the average. Subsequently, this strategy has been 
developed
in \cite{bologna} to include, among the others,  the REM \cite{derrida} 
and GREM \cite{grem},
and in \cite{franz-leone}, where finite connectivity models are considered.
In the present paper, we generalize the strategy of \cite{limterm} 
in a different direction. The class of models for which we provide a proof 
of
the existence of the thermodynamic limit embraces, for instance,
models where the spin degrees of freedom 
$\sigma_i$ have
several components, which are not necessarily two-valued Ising variables.
In the same way, we show how to treat the case where a Curie-Weiss 
interaction term is added to the mean field spin glass Hamiltonian.
Finally, the same results hold for a system composed of several replicas
(i.e., identical copies with the same disorder realization) of 
the mean field spin glass model, where replicas are coupled together by an
interaction term, which depends on their mutual overlaps.
In all of these cases, when the method of \cite{limterm} is naively 
applied, 
there appear terms which spoil the simple subadditivity argument
which works for the SK and $p$-spin models.
The main purpose of this paper is to show that the effect of these 
potentially
dangerous terms can be eliminated, by suitably 
decomposing the configuration space. This idea was introduced by Michel 
Talagrand in \cite{T}, and developed in a series of important applications.
Among these, Talagrand proposed a very interesting generalization of the 
broken replica bounds \cite{G} to the case of systems made of two coupled replicas. 

The organization of the paper is as follows. In Section 2, 
we introduce some basic definitions concerning mean field spin glass 
models.
In Section 3, we  state the main results of the paper, and we give some
physically meaningful examples of models to which they can be applied.
Section 4 contains the proof of the results. Finally, Section 
5 is dedicated to conclusions.

\section{Basic definitions}

In this Section, we recall some basic definitions concerning mean field
spin glasses, without making reference to any specific model.

The generic configuration $\sigma$ of the system is defined by $N$ spin 
degrees of freedom $\sigma_1,\sigma_2,\ldots$.
We suppose each $\sigma_i$ to belong to a set $\cal S$$\in\mathbb{R}^n$,
$n\in {\mathbb N}$, 
equipped with an {\sl a priori} measure $\nu$. For instance, the case 
$\cal{S}=$$\{-1,+1\}$ and $\nu=1/2(\delta_{-1}+\delta_{+1})$ corresponds 
to the usual Ising two-valued variables.
The Hamiltonian of the model,
$H_N(\sigma,J),$
depends on the spin configuration, on the system size $N$ and on some 
quenched disorder, which we denote as $J$. Of course, the Hamiltonian
can also depend on some additional external fields, e.g., on the 
magnetic field $h$. The mean field character of the model consists in the 
condition that, if two configurations $\sigma$ and $\sigma'$ are connected
by a permutation of the site indices, 
the random variables $H_N(\sigma,J)$ and $H_N(\sigma',J)$ have the same 
distribution.

For a given inverse temperature $\beta$, we introduce the disorder 
dependent
partition function $Z_{N}(\beta,J)$, 
the quenched average of the free energy per site
$f_{N}(\beta)$, and the Boltzmann-Gibbs state 
$\omega_J$, according to the definitions
\begin{eqnarray}\label{Z}
&&Z_N(\beta,J)=\int_{{\cal S}^N} d\nu(\sigma_1)\ldots d\nu(\sigma_N)
\exp(-\beta H_N(\sigma,J)),\\
\label{f}
&&-\beta f_N(\beta)=N^{-1} E\log Z_N(\beta,J),\\
\label{state}
&&\omega_{J}(A)=Z_N(\beta,J)^{-1}\int_{{\cal S}^N} 
d\nu(\sigma_1)\ldots d\nu(\sigma_N)
A(\sigma)\exp(-\beta H_N(\sigma,J)), 
\end{eqnarray}
where $A$ is a generic function of the $\sigma$'s.

Let us now introduce the important concept of replicas. 
Consider a generic number $n$ of independent copies
of the system, characterized by the spin
variables $\sigma^{(1)}_i$, $\sigma^{(2)}_i$, $\dots$,
distributed according to the product of Boltzmann-Gibbs states
$$\Omega_J=\omega^{(1)}_J \omega^{(2)}_J \dots\omega^{(n)}_J,$$
where each $\omega^{(\alpha)}_J$ acts on the corresponding set of
$\sigma^{(\alpha)}_i$'s, and all replicas are subject to the {\sl
same} sample $J$ of the quenched disorder. 
For a generic smooth function $F$ of the configuration of $s$ replicas, we
define the $\langle.\rangle$ averages as
$$\langle F(\sigma^{(1)},\sigma^{(2)},\dots)\rangle=E\Omega_J\bigl(F(
\sigma^{(1)},\sigma^{(2)},\dots)\bigr),$$
where the Boltzmann-Gibbs averages $\Omega_J$ acts on the replicated 
$\sigma$
variables, and $E$ is the average with respect to the disorder $J$.

\section{The existence of the thermodynamic limit}

The main object of interest in the theory is the quenched free 
energy $f_N(\beta)$. First of all,
one would like  to prove that it admits a well defined limit, for 
$N\to\infty$,
independently from the explicit calculation of the limit itself.

We restrict our analysis to the case of Gaussian models, i.e., models 
for which 
the  $H_N(\sigma,J)$ are (correlated) Gaussian random variables. Of course,
these random variables are 
fully characterized by their mean values 
$$
b_N(\sigma)=E H_N(\sigma,J)
$$
and covariance matrix
$$c_N(\sigma,\sigma')=E(H_N(\sigma,J)H_N(\sigma',J))-
E(H_N(\sigma,J))E(H_N(\sigma',J)).$$
In order to 
prove the existence of the thermodynamic limit,
we suppose that the following conditions are satisfied.
First of all, we require that
\begin{eqnarray}
\label{media}
\frac {b_N(\sigma)}N= g(m^{(1)}_N(\sigma),\ldots,m^{(k)}_N(\sigma))+
O(N^{-1}).
\end{eqnarray}
Here, $k\in {\mathbb N}$,   $g$ is a smooth function of class
$C^1$ and 
the $m^{(i)}_N(\sigma)$ are bounded functions,  with $N m^{(i)}_N$ 
additive in the system size. In other words,
\begin{equation}
\label{prop1}
|m^{(i)}_N(\sigma)|\le M\hspace{1 cm} \forall i,N,\sigma
\end{equation} 
and
\begin{eqnarray}
\label{prop2}
N m^{(i)}_N(\sigma)={N_1} m^{(i)}_{N_1}(\sigma^{(1)})+
{N_2} m^{(i)}_{N_2}(\sigma^{(2)}),
\end{eqnarray}
if $N=N_1+N_2$ and if the configuration $\sigma$ can be decomposed as
$$\sigma=(\sigma^{(1)}_1,\ldots,\sigma^{(1)}_{N_1},
\sigma^{(2)}_1,\ldots,\sigma^{(2)}_{N_2}).$$
As for the covariance matrix, we require that
\begin{eqnarray}
\label{covarianza}
\frac{c_N(\sigma,\sigma')}{N}
=f(Q^{(1)}_N(\sigma,\sigma'),\ldots,Q^{(k)}_N(\sigma,\sigma'))+
O(N^{-1}),
\end{eqnarray}
where $f$ is a convex function with continuous derivatives.
The variables  $Q^{(i)}_N$ must satisfy properties analogous to 
(\ref{prop1})-(\ref{prop2}), i.e.,
\begin{equation}
\label{prop3}
|Q^{(i)}_N(\sigma,\sigma)|\le M\hspace{1 cm} \forall i,N,\sigma
\end{equation} 
and
\begin{eqnarray}
\label{prop4}
N Q^{(i)}_N(\sigma,\sigma')={N_1} Q^{(i)}_{N_1}(\sigma^{(1)},\sigma'^{(1)})
+{N_2} Q^{(i)}_{N_2}(\sigma^{(2)},\sigma'^{(2)}).
\end{eqnarray}

It is interesting to notice that the models considered in \cite{limterm}
have the additional properties that $c_N(\sigma,\sigma)$ does not depend
on the configuration $\sigma$, and that $g$ is a linear function.

Now, we can state our result:
\begin{theorem}
\label{teorema1}
If conditions (\ref{media}) to (\ref{prop4}) are satisfied, then the 
thermodynamic limit of the quenched free energy exists:
\begin{equation}
\label{convmedia}
\lim_{N\to\infty}-\frac1{N\beta} E\ln Z_N(\beta)=f(\beta).
\end{equation}
Moreover, the disorder dependent free energy converges almost surely, 
with respect to the disorder realization:
\begin{equation}
\lim_{N\to\infty}-\frac1{N\beta} \ln 
Z_N(\beta)=f(\beta)\hspace{1cm}J-almost 
\;surely,
\end{equation}
and its disorder fluctuations can be estimated as
\begin{equation}
\label{deviazioni}
P\left(\left|-\frac1{N\beta}\ln Z_N(\beta)-f_N(\beta)\right|\ge 
u\right)\le 
2\exp\left(- \frac{N u^2}{2 L}\right),
\end{equation}
where 
\begin{equation}
\label{L}
L=\max_{|x_i|\le M\;\forall i}|f(x_1,\ldots,x_k)|,
\end{equation}
and $M$ is the same  constant as in (\ref{prop3}).
\end{theorem}

{\bf Remark} As explained in \cite{limterm}, from Eqs. 
(\ref{convmedia})-(\ref{deviazioni})
follows also  the convergence, both under quenched average and $J$-almost 
surely, of the ground state energy per site.

Before we turn to the proof of the Theorem, we give a few examples of 
physically meaningful systems to which it applies.

1. The SK model with non-Ising type spins, defined as
\begin{equation}
\label{SK}
H_N(\sigma,J)=-\frac1{\sqrt N}\sum_{1\le i<j\le N} J_{ij} \sigma_i\sigma_j-
h\sum_{i=1}^N \sigma_i.
\end{equation}
Here, and in the following examples, the couplings $J_{ij}$ are 
independent 
identically
distributed Gaussian random variables, with mean zero and unit variance,
while $h$ is the magnetic field. As for the spin degrees of freedom, we
suppose that
$\sigma_i\in{\cal S}=[-a,a]$, while the measure $\nu$ on ${\cal S}$, 
which appears in the definition of the partition function, is arbitrary.
In this case, 
$$
\frac{b_N(\sigma)}N=-h\, m_N(\sigma)=-\frac hN\sum_{i=1}^N\sigma_i
$$ 
and conditions (\ref{media}) to (\ref{prop2}) are clearly satisfied,
since $|m_N(\sigma)|\le a$ and  the total magnetization $\sum_i\sigma_i$
is linear in the system size.
Of course, the function $g$ in (\ref{media}) is just $g(x)=-h\,x$.
As regards the covariance matrix, one finds easily
$$\frac{c_N(\sigma,\sigma')}N=\frac{q_{\sigma\sigma'}^2}2+
O(N^{-1}),$$
where 
$$q_{\sigma\sigma'}=\frac1N \sum_{i=1}^N\sigma_i\sigma'_i, \hspace{1cm}
|q_{\sigma\sigma'}|\le a^2,$$
is the overlap of the two configurations. 
Since $N q_{\sigma\sigma'}$ is additive and $f(x)=x^2/2$ is convex,
conditions (\ref{covarianza}) to (\ref{prop4}) are also satisfied.

2. The SK model with an additional Curie-Weiss 
interaction, defined as
$$H_N(\sigma,J)=-\frac1{\sqrt N}\sum_{1\le i<j\le N} J_{ij} 
\sigma_i\sigma_j-
\frac {J_0}N\sum_{i,j=1}^N\sigma_i\sigma_j-h\sum_{i=1}^N \sigma_i,$$
where  $J_0$ is 
a non random constant and, again, $\sigma_i\in[-a,a]$. This model 
can be obtained from the previous one, if one supposes that the 
Gaussian variables $J_{ij}$ in (\ref{SK}) have mean value $2J_0/\sqrt N$.
This case can be dealt with in analogy with the previous one, with the 
only difference that 
$$
\frac{b_N(\sigma)}N=-J_0\,m_N(\sigma)^2-h\, m_N(\sigma),
$$ 
so that $g(x)=-h\,x-J_0\,x^2$.

3. The multi-replica SK model, with coupled replicas. In this
case, the Hamiltonian depends on the configurations $\sigma^{(1)},\ldots,
\sigma^{(n)}$ of the $n$ replicas, which interact through a term depending 
on the mutual overlaps:
$$H_N(\sigma^{(1)},\ldots,\sigma^{(n)},J)=
-\frac1{\sqrt N}\sum_{1\le i<j\le N} J_{ij} (\sigma^{(1)}_i\sigma^{(1)}_j+
\ldots+\sigma^{(n)}_i\sigma^{(n)}_j)+
N g(\{q_{ab}\}),
$$
where $g$ is a smooth $C^1$ function of all the overlaps.
The check of properties (\ref{media}) to (\ref{prop4}) is trivial, and is 
left
to the reader.

4. The SK model with Heisenberg type interaction, defined
by the Hamiltonian
\begin{equation}
H_N(\sigma,J)=-\frac1{\sqrt N}\sum_{1\le i<j\le N} J_{ij} \vec \sigma_i
\,\vec\sigma_j-
\sum_{i=1}^N \vec h\,\vec\sigma_i,
\end{equation}
where $\vec \sigma_i$ has $n$ bounded components $\sigma_i^{(1)},\ldots,
\sigma_i^{(n)}$, and $\vec u\,\vec v$ denotes scalar
product in ${\mathbb R}^n$.
In this case, 
\begin{equation}
\label{covar}
\frac{c_N(\sigma,\sigma')}N=\frac12\sum_{a,b=1}^n
(q^{ab}_{\sigma\sigma'})^2+O(N^{-1}),
\end{equation}
where 
$$
q^{ab}_{\sigma\sigma'}=\frac1N\sum_{i=1}^N\sigma^{(a)}_i\sigma'^{(b)}_i.
$$

It is instructive to verify explicitly that, for these models, the 
method introduced in \cite{limterm} does not work, and requires an
extension.

\section{Proof of Theorem 1}

We start with the proof of (\ref{deviazioni}). Results of this kind were 
firstly obtained in \cite{CIS} in the general context of the norms of 
Gaussian sample functions, and later developed in \cite{P}, and \cite{LD}.
The book by Talagrand \cite{T} demonstrates at length the usefulness of 
this idea in the applications to mean field spin glass theory. 
For later convenience,
we give a selfcontained proof of the more general inequality
\begin{equation}
\label{+generale}
P\left(\left|-\frac1{N\beta}\ln Z^A_N(\beta)+\frac1{N\beta}E\ln 
Z^A_N(\beta)
\right|\ge u\right)\le 2\exp\left(-\frac{N u^2}{2L}\right),
\end{equation}
where 
\begin{eqnarray}\label{ZA}
Z^A_N(\beta,J)=\int_{A_N} d\nu(\sigma_1)\ldots d\nu(\sigma_N)
\exp(-\beta H_N(\sigma,J))
\end{eqnarray}
is a modified 
disorder-dependent partition function, with the sum over configurations
restricted to an arbitrary
nonrandom set $A_N$ in the configuration space ${\cal S}^N$.

The restriction to a subset of the space of configurations has been 
exploited also in \cite{T} in the case of  the p-spin models.

We rewrite the Gaussian variables $H_N(\sigma,J)$ as
\begin{equation}
\label{riscritto}
H_N(\sigma,J)=\xi_N(\sigma)+b_N(\sigma),
\end{equation}
where, of course, $\xi_N(\sigma)$ 
is a centered Gaussian random variable, and 
$$E(\xi_N(\sigma)\xi_N(\sigma'))=c_N(\sigma,\sigma').$$
Given $s\in {\mathbb R}$, we define
\begin{equation}
\varphi_N(t)=\ln E_1 G_N(t)=\ln E_1 \exp\left(s\beta^{-1} E_2 \ln 
Z^A_N(t)\right),
\end{equation}
where the interpolating parameter $t$ varies between $0$ and $1$, 
and $Z_N^A(t)$ is the auxiliary partition function
\begin{eqnarray}
Z^A_N(t)&=&Z^A_N(t,J_1,J_2,\beta)\\\nonumber
&=&\int_{A_N} d\tilde\nu(\sigma)
\exp(-\beta\sqrt t \xi^1_N(\sigma)-\beta\sqrt{1-t} \xi^2_N(\sigma)-
\beta b_N(\sigma)).
\end{eqnarray}
Here, $\xi^1_N(\sigma)$ and $\xi^2_N(\sigma)$ are two {\sl independent} 
copies 
of the random
variable $\xi_N(\sigma)$, with the same distribution, and $E_1,E_2$ denote 
average with respect to $\xi^1$ and $\xi^2$, respectively.
For simplicity of notation, we set $$d\tilde\nu(\sigma)=
d\nu(\sigma_1)\ldots d\nu(\sigma_N)$$
the {\sl a priori} measure on ${\cal S}^N$, and 
$$p_N(t,\sigma)=
\frac{\exp(-\beta\sqrt t \xi^1_N(\sigma)-\beta\sqrt{1-t} \xi^2_N(\sigma)-
\beta b_N(\sigma))}
{Z^A_N(t)}$$
the modified Boltzmann weight.

It is very simple to check that
\begin{equation}
\label{bordi}
\varphi_N(1)-\varphi_N(0)=\ln E\exp\,{s\beta^{-1}}\left(\ln 
Z^A_N(\beta)-E\ln Z^A_N(\beta)\right).
\end{equation}
As for the $t$ derivative of $\varphi_N(t)$, an application of the formula
\begin{equation}
\label{byparts}
E x_i F(\{x\})=\sum_j E(x_i x_j)E \partial_{x_j}F(\{x\})
\end{equation}
which holds for any family of Gaussian random variables $\{x_i\}$ and 
any smooth function $F$, gives
\begin{eqnarray}
\nonumber
\varphi_N'(t)&=&\frac{s^2}{2E_1 G_N(t)}E_1\left\{ G_N(t)
\int_{A_N\times A_N}
\hspace{-1cm}d\tilde\nu(\sigma) d\tilde\nu(\sigma')
c_N(\sigma,\sigma') E_2 p_N(t,\sigma)E_2 p_N(t,\sigma')\right\}.
\end{eqnarray}
Thanks to the bound (\ref{prop3}), one has
\begin{equation}
|\varphi_N'(t)|\le\frac{s^2}{2}\max_{\sigma,\sigma'}|c_N(\sigma,\sigma')|=
\frac{N s^2}{2}\max_{\sigma,\sigma'}|f(Q_N(\sigma,\sigma'))|=\frac{N 
s^2L}{2}.
\end{equation}
Therefore, using Eq. (\ref{bordi}) and the obvious inequality 
$$
e^{|x|}\le e^x+e^{-x},
$$
one finds
\begin{equation}
E\exp\left(N|s|\left|\frac{1}{N\beta}\ln Z^A_N(\beta)-\frac{1}{N\beta}
E\ln Z^A_N(\beta)\right|\right)\le2\exp\left(\frac{s^2NL}2\right).
\end{equation}
By Tchebyshev's inequality,
\begin{equation}
P\left(\left|\frac{1}{N\beta}\ln Z^A_N(\beta)-\frac{1}{N\beta}
E\ln Z^A_N(\beta)\right|\ge u\right)\le2\exp\left(-N|s|u+\frac{s^2NL}2
\right)
\end{equation}
and, choosing $|s|=u/L$, one finally obtains the estimate 
(\ref{+generale}).
$\Box$

Now, we can prove the main statements of Theorem 1, concerning the 
existence of the thermodynamic limit. For simplicity, we assume that 
$$
N^{-1} b_N(\sigma)=g(m_N(\sigma))+O(N^{-1}),
$$
and
$$
N^{-1}c_N(\sigma,\sigma')=f(Q_N(\sigma,\sigma'))+O(N^{-1}),
$$
corresponding  to the case 
$k=1$ in Eqs. (\ref{media}), (\ref{covarianza}), and that
$L=1$ in (\ref{L}).
The general case can be obtained as a simple extension.

First of all, we prove the existence of the limit along sequences of the 
type
$\{N_K\}=\{N_0 n^K\}$, with $n,N_0\in\mathbb N$. As in \cite{limterm}, 
the idea is to find a suitable interpolation between the original system, 
of
size $N_K$, and a system composed of $n$ non interacting subsystems, 
of size $N_{K-1}$ each. However, in the present case, it is also 
necessary to divide 
the configuration space into sets, such that $m_N(\sigma)$ and $Q_N(\sigma,
\sigma)$ are approximately constant within each set. The idea of 
restricting to the set of configurations with given overlap was introduced 
by Michel Talagrand \cite{T}, and exploited in a series of important 
applications. 
For any $0<\varepsilon<1$, we can write
\begin{equation}
Z_{N_K}(\beta)=\sum_{i,j=0}^{[1/\varepsilon]}Z_{N_K}^{(ij)}(\beta)\equiv
\sum_{i,j=0}^{[1/\varepsilon]}\int_{C_{ij}}d\tilde\nu(\sigma)
\exp(-\beta H_{N_K}(\sigma,J)),
\end{equation}
where 
\begin{equation}
C_{ij}=\{\sigma\in {\cal S}^{N_K}: i\varepsilon\le Q_{N_K}(\sigma,\sigma),
<(i+1)\varepsilon, j\varepsilon\le m_{N_K}(\sigma),
<(j+1)\varepsilon\}
\end{equation}
and $[x]$ denotes the integer part of $x$.
Since $N_K=n N_{K-1}$, we can divide the system into $n$ subsystems of 
$N_{K-1}$ spins each, and we denote the configuration of the 
$\ell$-th subsystem as $\sigma^{(\ell)},\;\ell=1,2,\ldots,n$. 
Of course, the following inequality holds
\begin{equation}
Z_{N_K}^{(ij)}(\beta)\ge \tilde Z_{N_K}^{(ij)}(\beta)=
\int_{\tilde C_{ij}}d\tilde\nu(\sigma)\exp(-\beta H_{N_K}(\sigma,J)),
\end{equation}
where 
\begin{eqnarray}
C_{ij}\supseteq\tilde C_{ij}&=&\{\sigma\in {\cal S}^{N_K}:i\varepsilon\le
 Q_{N_{K-1}}(\sigma^{(\ell)},\sigma^{(\ell)})
<(i+1)\varepsilon,\\\nonumber
&&\hspace{2cm}j\varepsilon\le m_{N_{K-1}}(\sigma^{(\ell)})
<(j+1)\varepsilon,\;\forall \ell\}.
\end{eqnarray}
Now, we introduce an interpolating parameter $0\le t\le1$, and the 
auxiliary
partition function
\begin{eqnarray}\nonumber
\tilde Z_{N_K}^{(ij)}(t,\beta)&=&\int_{\tilde C_{ij}}d\tilde\nu(\sigma)
\exp\beta\left(-\sqrt{t}\xi_{N_K}(\sigma)-t b_{N_K}(\sigma)-
\sqrt{1-t}\sum_{\ell=1}^n\xi^\ell_{N_{K-1}}
(\sigma^{(\ell)})\right.\\\nonumber
&&\left.-(1-t)\sum_{\ell=1}^n b_{N_{K-1}}(\sigma^{(\ell)})\right),
\end{eqnarray}
where the $\xi_N^\ell(\sigma)$ are $n$ {\sl independent} copies of the 
random
variable $\xi_N(\sigma)$.
Clearly, for the boundary values of the parameter $t$ one has
\begin{equation}
\label{bordo1}
-\frac1{N_K\beta}E\ln \tilde Z_{N_K}^{(ij)}(0,\beta)=-\frac1{N_{K-1}\beta}E
\ln Z_{N_{K-1}}^{(ij)}(\beta)
\end{equation}
and
\begin{equation}
\label{bordo0}
-\frac1{N_K\beta}E\ln \tilde Z_{N_K}^{(ij)}(1,\beta)=
-\frac1{N_K\beta}E\ln \tilde Z_{N_K}^{(ij)}(\beta)\ge-\frac1{N_K\beta}
E\ln Z_{N_K}^{(ij)}(\beta).
\end{equation}
As regards the $t$ derivative, we apply the integration by parts 
formula (\ref{byparts}) and, 
recalling that the random variables $\xi_N^\ell(\sigma)$ are statistically
independent for different $\ell$, we find after some long but 
straightforward
computations,
\begin{eqnarray}
\label{primo}
\nonumber
&&-\frac{d}{dt}\frac1{N_K\beta}E\ln \tilde Z_{N_K}^{(ij)}(t,\beta)=\\
&&\hspace{1cm}-\frac{\beta}2\Med{f(Q_{N_K}(\sigma,\sigma))-
\frac1n\sum_{\ell=1}^nf(Q_{N_{K-1}}(\sigma^{(\ell)},\sigma^{(\ell)}))}
\\
\label{secondo}
&&\hspace{1cm}+
\frac{\beta}2\Med{f(Q_{N_K}(\sigma,\sigma'))-
\frac1n\sum_{\ell=1}^nf(Q_{N_{K-1}}(\sigma^{(\ell)},\sigma'^{(\ell)}))}
\\
\label{terzo}
&&\hspace{1cm}+\Med{g(m_{N_K}(\sigma))-\frac1n\sum_{\ell=1}^ng(m_{N_{K-1}}(
\sigma^{(\ell)}))}+O\left(\frac1{N_K}\right),
\end{eqnarray}
where the averages are, of course, restricted to configurations belonging
to $\tilde C_{ij}$.
Since $Q_{N_K}$ is a convex combination of the $Q_{N_{K-1}}$ and 
$f$ is a convex function, the term (\ref{secondo}) is not positive.
On the other hand, since $f$ is a 
function of class $C^1$, and  $Q_{N_{K-1}}(\sigma^{(\ell)},
\sigma^{(\ell)})$ and are constrained to belong
to the interval $[i\varepsilon,(i+1)\varepsilon)$ for each $\ell$, the term
(\ref{primo}) is of order $\varepsilon$. The same holds for the term 
(\ref{terzo}).
This implies that, for $K$ large enough,
\begin{eqnarray}
-\frac{d}{dt}\frac1{N_K\beta}E\ln \tilde Z_{N_K}^{(ij)}(t,\beta)
\le C\varepsilon,
\end{eqnarray}
for some positive constant $C$ independent of $N$.
Recalling Eqs. (\ref{bordo1}), (\ref{bordo0}), this means that
\begin{equation}
\label{j}
-\frac1{N_K\beta}E\ln Z_{N_K}^{(ij)}(\beta)+
\frac1{N_{K-1}\beta}E\ln Z_{N_{K-1}}^{(ij)}(\beta)\le C\varepsilon.
\end{equation}
Now, we want to turn this inequality, which involves disorder averages, 
into 
an inequality valid $J-$almost everywhere. To this purpose,  
we choose $\varepsilon=N_K^{-1/4}$ and we observe that, thanks to 
the estimate (\ref{+generale}),
\begin{eqnarray}
P\left(-\frac1{N_K\beta}\ln Z_{N_K}^{(ij)}(\beta)\ge
-\frac1{N_K\beta}E\ln Z_{N_K}^{(ij)}(\beta)+C\varepsilon\right)\le 2 
\exp\left(
-\frac{\sqrt{N_K}C^2}2\right)
\end{eqnarray}
and
\begin{eqnarray}
P\left(-\frac1{N_{K-1}\beta}\ln Z_{N_{K-1}}^{(ij)}(\beta)\le
-\frac1{N_{K-1}\beta}E\ln Z_{N_{K-1}}^{(ij)}(\beta)-
C\varepsilon\right)\le 2 \exp\left(-\frac{\sqrt{N_K}C^2}{2n}\right).
\end{eqnarray}
Therefore, with probability 
$P\ge 1-4\sqrt{N_K}\exp\left(-\frac{\sqrt{N_K}C^2}{2n}\right)$, one has
\begin{eqnarray}
\label{borelcantelli}
-\frac1{N_K\beta}\ln Z_{N_K}^{(ij)}(\beta)\le
-\frac1{N_{K-1}\beta}\ln 
Z_{N_{K-1}}^{(ij)}(\beta)+3CN_K^{-1/4}\hspace{.5cm}
\forall i,j=0,\ldots,[N_K^{1/4}].
\end{eqnarray}
Since the probability of the complementary event is summable in $K$, it 
follows from Borel-Cantelli lemma \cite{shiri} that inequality 
(\ref{borelcantelli}) holds $J$-almost surely, for $K$ large enough.
As a consequence, one obtains
\begin{eqnarray}
Z_{N_K}(\beta)&=&\sum_{i,j=0}^{[N_K^{1/4}]}Z_{N_K}^{(ij)}(\beta)\ge
e^{-3\beta 
CN_K^{3/4}}\sum_{i,j=0}^{[N_K^{1/4}]}\left(Z_{N_{K-1}}^{(ij)}(\beta)
\right)^n\\
\nonumber
&&\ge e^{-3\beta CN_K^{3/4}}N_K^{(1-n)/2}
\left(\sum_{i,j=0}^{[N_K^{1/4}]}Z_{N_{K-1}}^{(ij)}(\beta)\right)^n=
\frac{e^{-3\beta CN_K^{3/4}}}{N_K^{(n-1)/2}}Z_{N_{K-1}}^n(\beta).
\end{eqnarray}
Here, we have used the property
\begin{equation}
\sum_{i=1}^k x_i^n\ge k^{1-n}\left(\sum_{i=1}^k x_i\right)^n,
\end{equation}
which holds if $x_i\ge 0$, thanks to the convexity of the function $x^n$.
By taking the logarithm and dropping terms of lower order in $N_K$, one has
\begin{eqnarray}
\label{inessential}
-\frac{1}{N_K\beta}\ln Z_{N_K}(\beta)\le -\frac{1}{N_{K-1}\beta}\ln 
Z_{N_{K-1}}(\beta)
+3C N_K^{-1/4}\hspace{.5cm}J-a.s.,
\end{eqnarray}
for $K$ large enough. Notice that, with respect to (\ref{j}), 
the above inequality involves the original free energy, where the sum over 
over configuration has no restrictions.
From (\ref{inessential}), it follows that the thermodynamic limit exists, 
$J$-almost surely, the term $N_K^{-1/4}$ being inessential. On the other 
hand, the exponential estimate (\ref{deviazioni}), together with 
Borel-Cantelli Lemma, implies that the limit has a non
random value $f(\beta)$, for almost every disorder realization $J$.

Once the almost sure convergence is proved, the convergence of the 
quenched 
average can be obtained easily, provided that 
the probability that $1/N\ln Z_N$ assumes
large values is sufficiently small. For instance, 
one has the following criterion \cite{picco}:
given random variables $X_K$ and $X$, if $X_K\longrightarrow X$ almost 
surely 
for $K\to\infty$, and if
\begin{equation}
\label{sup}
\lim_{\lambda\to\infty}\sup_K E\left(|X_K|\,{\cal X}(\{|X_K|\ge \lambda\})
\right)=0,
\end{equation}
where ${\cal X}(A)$ denotes the characteristic function of the set $A$, 
then
$$
E X_K\to E X.
$$
In the present case, $X_K=-(1/{N_K\beta}) \ln Z_{N_K}(\beta)$, $X=f(\beta)$,
and the condition (\ref{sup}) can be easily checked, by employing the
exponential bound (\ref{deviazioni}).

In conclusion, we have proved almost sure convergence for the free 
energy, and convergence of its quenched average, for any 
subsequence of the form
$\{N_0 n^K\}$. It is not difficult  to show, by standard methods, that 
this 
implies convergence along any increasing subsequence $\{N_K\}$, and the 
uniqueness of the limit. $\Box$

\section{Conclusions}

In this paper, we have extended the class of mean field spin glass models
for which the existence of the thermodynamic limit can be proved, 
independently
of an explicit calculation of the limit itself. Essentially, with respect
to the models considered in \cite{limterm}, one abandons the assumption
that the variance of the Hamiltonian is independent of the configuration,
and that its mean value is additive in the size of the system. 
Some of the hypotheses of the Theorem, like the uniform
bounds (\ref{prop1}), (\ref{prop3}), are required only for technical 
reasons,
but can be dispensed with, at the expense of some extra work. 
On the other hand, the condition of convexity for the covariance 
function is essential, so that our result does not extend directly, 
for instance, to the $p$-spin model, with $p$ odd.

\vspace{.5cm}
{\bf Acknowledgments}

We gratefully acknowledge useful conversations with
Pierluigi Contucci, Sandro Graffi, Giorgio Parisi and Michel Talagrand.

This work was supported in part by MIUR 
(Italian Minister of Instruction, University and Research), 
and by INFN (Italian National Institute for Nuclear Physics).

\end{document}